\documentclass[nofootinbib,longbibliography]{revtex4-2}

\usepackage{amsmath}
\usepackage{amsfonts}
\usepackage{amssymb}
\usepackage{bbm}
\usepackage{graphicx}
\DeclareMathOperator*{\argmax}{argmax}

\DeclareMathOperator{\Tr}{Tr}

\begin{document}

\title{Towards fully bayesian analyses in Lattice QCD}
\author{Julien Frison}
\email[]{julien.frison@desy.de}
\affiliation{John von Neumann-Institut f\"ur Computing NIC, Deutsches Elektronen-Synchrotron DESY, Platanenallee 6, 15738 Zeuthen, Germany}

\begin{abstract}
We present a promising method to learn physical parameters from a bayesian inference, using modern tools to replace both our traditional fits and the way errors are computed and propagated.
A few models are built as illustrations for a realistic case with Lattice QCD data, and appear to extract a lot of information with good stability.
We discuss the evaluation of these models with either a fully bayesian approach or information criteria, as well as the model-building challenges which remain to be solved.
\end{abstract}

\maketitle

\section{Introduction}

Bayesian inference offers a well-defined way to interpret statistically-distributed data and learn physical parameters, which can be applied with great flexibility to many kinds of models without needing any specific assumption.
It is usually opposed to frequentist methods, but is arguably superior in the sense that any frequentist statement can be reinterpreted as a bayesian statement with a hidden prior. The main drawback
of bayesian methods is their higher use of computing resources.
Some progress has been made in this regard and some powerful software is available, which is nowadays routinely used in many domains of the academic world and even in the industry, as one
of the tools of Machine Learning. This is
likely to improve even further in the future, joining the efforts of the statistician community  as well as
those of many other disciplines. In this paper, we
demonstrate some use of these generic tools in the case of the typical analyses of Lattice Quantum Field Theory (LQFT).

As of today, the state-of-the-art in LQFT consists in non-linear weighted (WLS) or generalised (GLS) least square, usually combined with some resampling techniques\cite{efron1992bootstrap}
such as non-overlapping block bootstrap.
The resampling is sometimes replaced by a linear propagation of errors\cite{Wolff:2003sm}\cite{Ramos:2018vgu}. Finally, pseudo-bayesian model averaging (pBMA) has become more and more popular\cite{Jay:2020jkz} and acts as a third layer.
The ambition of our work is to replace all that by a unified framework performing everything
in a single step with well-defined interpretation, a fully and explicitly bayesian analysis.

We will first present in Sec.~\ref{sec:fundamentals} our notations and some introduction to the fundamentals of bayesian analysis. The reader already familiar with bayesian
analysis might choose to skip this section, as well as Sec.~\ref{sec:basic-toy} where we consider some basic toy models to illustrate this.
In Sec,~\ref{sec:traditional} we come back to the traditional method, applied to actual LQFT data.
In Sec.~\ref{sec:models} we present bayesian models for this same data, then in Sec.~\ref{sec:IC} we consider how to select and average the models we build, and finally
in Sec.~\ref{sec:misspecified} we discuss the case of misspecified models.

\section{Fundamentals}
\label{sec:fundamentals}
\subsection{Notations}
Throughout this paper, we will consider a set of data $y$, made of $n$ samples of vectors $y_i$ of dimension $p$, and some model $M$ with parameters $a$, $\dim(a)$=k.
The empirical mean $\sum_i y_i/n$ will be noted $\bar{y}$.
What we call model is a function $a\mapsto M(a)$ where $M(a)$ is a distribution.
We will note $M^*$ the {\it true} model, i.e. the model which has been used to generate the data $y$, which we have no way to actually know
with a finite amount of data.
There might or not exist a {\it true} parameter $a^*$ such that $M(a^*)=M^*$. 
If there is we will call $M^*$ parametrizable by $M$.
If there is not, we will call the model $M$ misspecified.
In this paper we use a general notation $P(X|Y)$ for the conditional probability of $X$ knowing in advance some information $Y$, and similarly $E(\dots)$ for an expectation value or $V(\dots)$ for a variance.
A normal distribution with mean $\mu$ and standard error $\sigma$ will be noted ${\cal N}(\mu, \sigma)$.

\subsection{The Bayes formula}
While we do not have a direct access to $a^*$ we can express our knowledge in terms of the Bayes formula
\begin{equation}
\label{eq:bayes}
P(a|y,M) = \frac{P(y|a,M)P(a|M)}{P(y|M)} .
\end{equation}
The left hand side is called the posterior distribution, while the data-independent $P(a|M)$ is a semi-arbitrary distribution called prior. They are related through the likelihood $P(y|a,M)$ which
encodes the core information of the model, while the denominator $P(y|M)$ is called marginal distribution because it can be obtained  by marginalising (integrating) $a$ in the numerator. The marginal distribution
can usually be viewed as a constant normalisation which does not need to be computed in practice.

A prior is said to be flat if, before observing any data, all values of $a$ have the same probability. It is said
to be uninformative if it is reasonably close to being flat, compared to the amount of information present in the data, 
so that the behaviour of the posterior is dominated by the influence of the likelihood.

Another distribution we are going to use is the posterior predictive (PPD), built from the posterior as
\begin{equation}
P(y'|y,M) = \int P(y'|a,M) P(a|y,M) da .
\end{equation}
The existence of this PPD makes bayesian models a part of the family of {\it generative} machine-learning
models.

\subsection{Maximum likelihood and maximum a posteriori}
\label{sec:MLE}

We are going to use two closely related point estimators.

Frequentist analyses often use the maximum likelihood estimator (MLE), given by
\begin{equation}
  a_{MLE} = \argmax_a P(y|a,M) .
\end{equation}

A similar definition, more useful in our case, is the maximum a posteriori (MAP)
\begin{equation}
  a_{MAP} = \argmax_a \left[P(y|a,M)P(a|M)\right] ,
\end{equation}
which is the mode of the posterior distribution. This quantity does not give any information on the uncertainty of the parameters, but one can use the Hessian to build locally a Gaussian approximation,
which is what the $\delta$ and $\Gamma$ methods\cite{Wolff:2003sm} do.

This can be a dangerous object to manipulate, since the likelihood can in principle have local maxima in sharp peaks or funnels which do not actually
represent a large share of the volume of probability. When this happens, reparametrising the models appropriately can help because the MLE and MAP, 
unlike volumes of posterior probability, are not invariant through reparametrisation.

The MLE and the MAP agree when the prior is flat.
Interestingly, a flat prior is not a theoretically superior choice. Actually it has been demonstrated that the MLE (and therefore a flat prior) is a bad choice in large dimension\cite{stein1956}, 
in the sense that one can always find a more efficient estimator without sacrificing anything in the bias-covariance trade-off. 
And once again what {\it flat} means is not invariant through reparametrisation.

\subsection{Least square estimates}

The least square procedure consists in minimising a function $a\mapsto \sum_i||y_i-f(a)||$, or for the generalised least square:
\begin{equation}
\chi^2_C(a) = \sum_{i=1}^n \left[y_i-f(a)\right]^\dag C^{-1} \left[y_i-f(a)\right],
\end{equation}
where the $p\times p$ matrix $C$ is some input imposed a priori and kept fixed during the minimisation.

This corresponds to computing the MLE for the gaussian likelihood
\begin{equation}
P(y|a,M_C) \propto e^{-\chi^2_C(a)/2} \propto e^{-\frac{n}{2}\left[\bar{y}-f(a)\right]^\dag C^{-1} \left[\bar{y}-f(a)\right]},
\label{eq:gaussianLike}
\end{equation}
where we note $M_C$ a model which contains $C$ fixed as a part of its intrinsic properties but keeps $a$ free, so all the parameters on which the minimisation is performed act on the mean of the gaussian distribution.

Note that Eq.~\ref{eq:gaussianLike} can be written either in terms of $y$ ($n$ data points fitted with a covariance $C$)
or $\bar{y}$ ($1$ data point fitted with a covariance $C/n$), regardless of the true distribution of $y$.
This means that, once $C$ is chosen, enforcing a gaussian model for each $i$ separately or only for the mean makes no difference on
the posterior or the MLE. But this would impact the discussion of fit quality in Sec.~\ref{sec:IC}.
In the specific case where $C$ is chosen to be the empirical covariance of $y$, this relates to the Central Limit Theorem\cite{Billingsley}:
the asymptotic distribution of $\bar{y}$ depends only on the (auto)covariance of $y$ and does not depend on higher moments of $y$.
This is a property used by the $\Gamma$ method on primary observables\footnote{Eq.~(11) of \cite{Wolff:2003sm} is essentially Eq.~(27.20) of \cite{Billingsley}}.

In standard lattice analyses, $C$ is either set to the empirical covariance of the data, or only its diagonal part. However, even if the {\it true}
distribution were perfectly gaussian, we could not have access to its {\it true} covariance parameter with a finite amount of data. This is one of the issues we will treat in this paper.

A $L_2$ regularisation\cite{ridge} is sometimes added, and absorbed into an augmented $\chi^2$.
It is often improperly called a bayesian prior, because it is equivalent to the MAP of bayesian model with a gaussian prior.
Such a regularisation still allows to write the function to minimise as a least square and use the Levenberg-Marquardt algorithm\cite{levenberg1944method}, while
a MAP with arbitrary priors might require instead more generic minimisers, which are typically slower and less stable. This
limitation becomes less important nowadays with the development of powerful optimisation methods for machine learning.

No estimate of the error is included in the least square method itself, which is a point estimate, and traditional lattice methods
need to bring an extra layer of analysis to take them into account.

\subsection{Sampling with PyMC}

To obtain a fully bayesian analysis, one would need to obtain not just a point estimate but the full posterior distribution of all parameters $a$.
There are a few cases where this can be written analytically, but most models do not allow that. However, for any value of $a$ we
are able to compute the associated likelihood $P(y|a)$, and therefore up to a constant the posterior distribution $P(a|y)$. Monte-Carlo techniques
allow to sample $a$ according to this probability, regardless of what the constant is.

A few ``standard'' programs and packages are on the market for this kind of bayesian inference.
We chose to base this work on PyMC\cite{PyMC}, a Python package which allows to build and fit models in a few lines.
These tools are typically based on an algorithm developed by the lattice community: the Hybrid Monte-Carlo\cite{DUANE1987216}.

The parameters $a$ are then analogous to a gauge configuration, while the log-likelihood of our model is analogous to a lattice action and the marginal likelihood is a partition function.
As long as $a$ is made of continuous parameters, values of $a$ can be sampled very efficiently according to the posterior probability,
even in large dimension.
Indeed, even for the most complicated models we will present and for models where minimisation would be
very difficult, the dimensionality is much lower that
what we typically encounter in the lattice computations the HMC was made for. 
The forces are computed through automatic differentiation with Theano\cite{theano}.
The set of all values of $a$ along the Monte-Carlo history is called a trace. A trace usually contains several Monte-Carlo chains in parallel, as a probe for ergodicity.

The results shown in this paper are obtained on a couple of cores of an Intel Xeon Gold 6130 CPU.
It could also realistically be run on a laptop, while for the most complicated models a GPU would have been a particularly interesting choice, already supported by those libraries.
All the Jupyter notebooks used to generate the results and figures of this paper are available on \cite{zenodo}.

\section{An example of traditional bootstrapped fit}
\label{sec:traditional}

Before plunging into the application of bayesian methods, let us set up a reference implementation of a traditional
bootstrapped least square fit, the way it is usually done as of today. For consistency, we present results obtained in a
PyMC implementation. However we obtained qualitatively similar results with a well-tested GSL-based C++ code, so the technical choice of
minimiser does not seem to be crucial.
Here and throughout this paper our data is a real-life pion correlator from 1009 configurations of the CLS ensemble H101.
We stick to this particular example for consistency but our arguments apply to any kind of fit.

This example will illustrate (only) one of the many ways a least square fit can fail. Indeed four things can in principle
happen with non-linear functions. A first type of failure would occur in the case where the MLE is peaked at a
global maximum whose neighbourhood only contains a small volume of probability. 
A second type, almost as annoying from the theoretical point of view, occurs when local maxima of the likelihood exist, 
relatively close to the starting point of the minimiser.
A third type would be a mere problem of numerical instability, depending on initial guesses, parametrisation and regulators.
Lastly, a fourth type would be the case of a fit which gives reasonable results but fails to provide a reliable goodness-of-fit.
The example of this section will suffer mostly from type two failure.

We start by enlarging our data with a new dimension representing the bootstrap:
\begin{equation}
z_{b,i} = y_{\beta_b(i)}
\end{equation}
where the functions $\beta_b$ are drawn randomly in $[1,n]^{[1,n]}$ for each $b$ (we draw $N_b=200$ of them).

\begin{figure}[ht]
      \hfill
\begin{center}
      \includegraphics[width=0.4\textwidth]{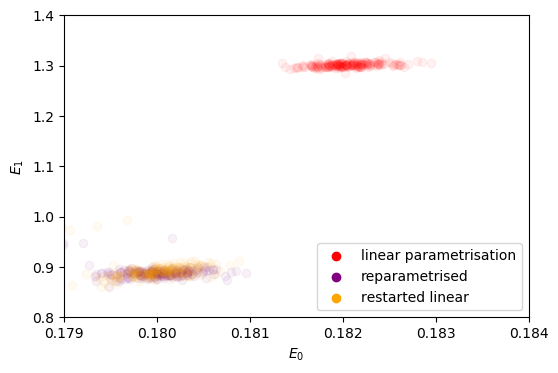}
      \hspace{0.1\textwidth}
      \includegraphics[width=0.4\textwidth]{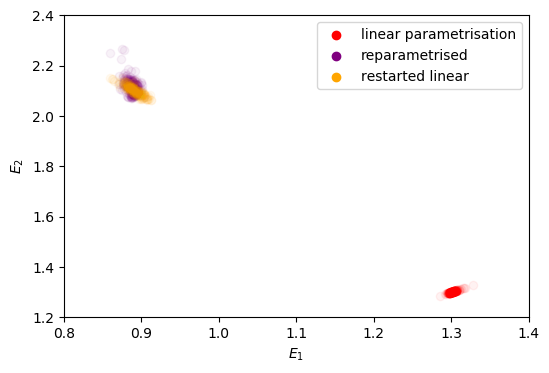}
\end{center}
      \hfill
\caption{We show the correlation between the energies of two states in a three-state uncorrelated least-square fit.
Each point represents a bootstrap sample. The restarted (yellow) results correspond to the linear (red) model
but minimised from a fine-tuned initial guess obtained from the more stable reparametrised (purple) model.
Results can change substantially depending on details of the implementation and choice of hyperparameters.
}
\label{fig:traditional}
\end{figure}

Using Eq.~\ref{eq:gaussianLike} with $a=(v,E)$, we write our fitting formula\footnote{Our PyMC model also
includes a hidden reparametrisation ordering the energies. It could also be sorted afterwards as in the C++ code}
\begin{eqnarray}
P(z|v,E) &\propto& \prod_{b=1}^{N_b} e^{-\frac{n}{2}\sum_t\left[\bar{z_b}(t)-f_t(v_b,E_b)\right]^2/\sigma^2(t)}\\
f_t(v_b,E_b) &=& \sum_{j=1}^{N_{\rm exp}} v_{bj} e^{-E_{bj} t}.
\end{eqnarray}
The MLE of both $E$ and $v$ are now a $N_b\times N_{\rm exp}$ matrix.
We will call this the linear model (linear in $v$), and for $N_{\rm exp}=3$ we can see in 
Fig.~\ref{fig:traditional} that this gives very suspicious results.
Indeed the bootstrap claims to know $E_1$ and $E_2$ with good precision but those are degenerate, and at the same time the
variance of $v$ is very large.

In this case, a closer look allows to understand why our results got attracted towards this degeneracy: the line of degeneracy
is a long flat direction in the likelihood because any positive contribution in one component could be compensated by a
negative component in the other. One can even solve analytically for the linear variables $v$ for each $E$, and see that
in the degenerate limit the linear system to solve becomes singular, so that the local minima are sent to large values of $v$.
This degeneracy line might be very thin and represent only a small integrated probability, 
but the minimiser is not sensitive to that: it only looks at infinitesimally small volumes.

We can then propose a reparametrisation which mostly solves our problem, despite leaving a non-negligible amount of outliers:
\begin{equation}
g_t(w_b,E_b) = \sum_{j=1}^{N_{\rm exp}} w_{bj}^2 e^{-E_{bj} t}.
\end{equation}
This reparametrisation is equivalent to obtaining the MAP from a {\it half-flat} prior (times a jacobian) requiring $v$ to be positive,
but in this specific case fortunately we could avoid adding a more informative prior.
As we can see on Fig.~\ref{fig:traditional}, there is some strong inconsistency in terms of bootstrap error between this
reparametrised model and the previous linear model.
However the $\chi^2$ per degree of freedom is much better, and the results make more sense.
If we now repeat the minimisation of the linear model using as a starting point bootstrap-by-bootstrap the very precise guess
given by the result of the reparametrised model, the results dramatically improve (some marginal tension with the bayesian result will remain).

The conclusion of all this is that bootstrapped fits cannot be used as a black box for complicated non-linear models.
The convergence of the minimiser does not by itself guarantee the reliability of the boostrap errors. The results have
to be evaluated on a case-by-case basis and lead to ad-hoc solutions which can be difficult to trust.
The bayesian inference, applied to the same model in Sec.~\ref{sec:uncorr-model}, will be much less sensitive
to initial values and (implicit or explicit) priors.

\section{A few very basic toy models}
\label{sec:basic-toy}

\subsection{One-dimensional gaussian}
\label{sec:one-dim-gauss}

Let us start with the case $p=1$ of a gaussian model. In this case the $y_i$ are just real numbers and $a=(\mu,\sigma^2)$. $y_i$ could for instance
be values of the plaquette for each configuration $i$.

Let us impose flat priors on $\mu$, $\sigma^2$. In principle this can only be done as taking some family of prior towards a non-informative limit, and
what we have in mind in particular here is a wider and wider gaussian for $\mu$ and an improper distribution $\Gamma(0,0)$ (an object we will present later) for $\sigma^2$.
In practice the limits usually do not need to be written and in PyMC we can simply set
\begin{eqnarray}
\log P(\mu|M) &=& 0\\
\log P(\sigma^2|M) &=& 0.
\end{eqnarray}

The likelihood which defines our model is
\begin{equation}
P(y|M,\mu,\sigma^2) = \prod_{i=1}^n \frac{1}{\sigma\sqrt{2\pi}} e^{-\frac{1}{2}\left(y_i-\mu\right)^2}.
\end{equation}

This is a specific case where the posterior distribution (Eq.~\ref{eq:bayes}) can be computed analytically, and it is represented by the well-known Normal-Inverse-Gamma
distribution as
\begin{eqnarray}
P(\mu,\sigma^2|y,M) &=& NIG(\bar{y},\lambda,\alpha,\beta) \quad\mathrm{where}\\
&& \lambda = n\\
&& \alpha = n/2\\
&& \beta = \frac{1}{2}\sum_{i=1}^n (y_i-\bar{y})^2\\
NIG(\bar{y},\lambda,\alpha,\beta) &\equiv& \frac{\sqrt{\lambda}}{\sqrt{2\pi\sigma^2}}\frac{\beta^\alpha}{\Gamma(\alpha)}
    \left(\frac{1}{\sigma^2}\right)^{\alpha+1}\exp\left(-\frac{2\beta+\lambda(\mu-\bar{y})^2}{2\sigma^2}\right).
\end{eqnarray}

This is simply a gaussian in $\mu$, with average and MAP $\bar{y}$. In the $\sigma$ direction things are more complicated, and at this stage
we do not need to fully understand the behaviour of this distribution, but we can note that we know its mean and mode:
\begin{eqnarray}
E(\sigma^2) &=& \frac{\beta}{\alpha-1} \to \frac{1}{n}\sum_{i=1}^n (y_i-\bar{y})^2\\
\sigma^2_{MLE} &=& \frac{\beta}{\alpha + 3/2} \to \frac{1}{n}\sum_{i=1}^n (y_i-\bar{y})^2.
\end{eqnarray}

This posterior distribution is the final result of our bayesian inference. Not only we have information of the MLE $(\bar{y},\sigma^2_{MLE})$, including the $\bar{y}$ that we would get
as the result on a least square fit at fixed $\sigma^2$, but we are also able to provide confidence level intervals for the values of those parameters and have information about
their correlations one with the other.

\subsection{The trivial model}
\label{sec:trivial}

Let us come back to the case where $\sigma$ is fixed, like for a $\chi^2$ fit, but with now $p\not=1$. More specifically, let us first have a look
at the case with one fit parameter for the mean of each data component, $k=p$ and $f_i(a)=a_i$, 
with the likelihood of Eq.~\ref{eq:gaussianLike} and flat priors on $a$. This is what we will call the trivial model,
because there are zero degrees of freedom and the MLE is just a repetition of the data.

Here the posterior distribution of $a$ is actually a multivariate gaussian, with mean $\bar{y}$ and covariance $C/n$, where $C$ is the (arbitrary) matrix
which entered the likelihood. In the large $n$ limit $a$ will be perfectly known, regardless of the size of the fluctuations of $y_i$ allowed
by the model at the level of individual data points.

This model will become interesting again when talking about information criteria, or when applying cuts to fitting intervals.

\subsection{Reinterpreting least-square methods}
\label{sec:reinterpret}

In Eq.~\ref{eq:gaussianLike}, writing the GLS as a gaussian likelihood, we used the parameters $a$ to describe its means, but the covariance $C$ was considered to be {\it known} in advance,
i.e. before performing the bayesian inference or the minimisation.
However, in reality we have an imperfect knowledge of $C$, which is a model parameter estimated from the same data as the other parameters in $a$.
Forcing $C$ to be considered as perfectly known in our bayesian model is equivalent to putting a delta prior
\begin{equation}
  P(C|M) = \delta(C- \langle \left(y-\bar y\right)\left(y - \bar y\right)^T\rangle )
\end{equation}
This is not a very bayesian way of thinking: First, in a bayesian analysis the model, the prior and the data should ideally be three distinct things. 
Secondly, the $\delta$ function is way too informative as a prior: once the prior is set, the bayesian inference is stuck with an {\it incorrect} guess
and no amount of new data put into it could make $C$ change. Of course if we obtain some new data we will change the prior, but this would happen
outside of the nice statistical framework we set up, so the probabilistic interpretation of our results is affected.

In the particular case of the trivial model the empirical data covariance is almost the MLE of $C$ (modulo a $1/n$ bias).
There, freezing $C$ can only affect our estimation of error bars. In the general case
however this is not guaranteed. In particular for correlated fits in large dimension, the data covariance can easily be a non positive-definite matrix within machine precision.
If the dimension is larger than the number of samples, basic algebra can even {\it prove} that it is singular, so that a model using its pseudo-inverse leads to zero likelihood.

\section{Bayesian models}
\label{sec:models}
\subsection{The Wishart distribution}
\label{sec:wishart}
As we turn towards a fully bayesian analysis where the prior is relaxed, we have to decide what its functional form should be. It turns out that the Wishart distribution
\begin{equation}
  {\cal W}(C^{-1}|V,\nu) = \frac{|C^{-1}|^{(\nu-p-1)/2}e^{-\Tr (V^{-1}C^{-1})/2}}{2^{\frac{\nu p}{2}}|V|^{\nu/2}\Gamma_p(\frac{\nu}{2})}
\end{equation}
is a particularly interesting choice,
with a clear interpretation and simplified computations. Here we note $p$ the dimension of the scale matrix $V$ and $\Gamma_p$ is the multivariate $\Gamma$ function (not to confuse with
the $\Gamma$ distribution).

Indeed, this is the conjugate prior for the gaussian likelihood with known mean. This means that if we start with a Wishart prior we obtain a Wishart posterior. If we start from a very
uninformative Wishart prior ($\nu\sim p$) and add some data, we get a more informative Wishart distribution, where $\nu'=\nu+n$ is increased by the amount of data points $n$ used in this round of inference and $\nu' V'$
approaches the empirical inverse covariance $C^{-1}$ computed on this data.

In the scalar case $p=1$ (such as Sec.~\ref{sec:one-dim-gauss}), the Wishart distribution is called the $\Gamma$ distribution, and similarly its expression is
\begin{equation}
  \Gamma(x|\alpha,\beta) = \frac{\beta^\alpha}{\Gamma(\alpha)}x^{\alpha-1}e^{-\beta x}.
\label{eq:gamma}
\end{equation}
It is a generalisation of the $\chi^2$ distribution for non-integer degrees of freedom (this role is played by $\alpha$ or $\nu$).

Finally, the posterior predictive distribution is known as well, in the case of a gaussian likelihood with known mean and a Wishart prior: it is given by the multivariate Student-t distribution
\begin{equation}
  t_{\nu+n-p+1}\left(y\mid\mu,\frac{\left(V^{-1}+\sum_{i=1}^n(y_i-\mu)(y_i-\mu)^T\right)^{-1}}{\nu+n-p+1}\right).
\label{eq:PPD-t}
\end{equation}

For $n$ very large, the Student-t distribution converges to a gaussian, and this provides an asymptotic justification to the model of Sec.~\ref{sec:reinterpret}.
However two differences appear: First, the scale matrix is computed from the fit parameter $\mu$ rather than from the data average $\langle y\rangle$.
Secondly, for finite values of $n$ this distribution has a wider tail.
There, a weaker penalisation of outliers makes sense since $\langle(y_i-\mu)(y_i-\mu)^T\rangle$ is not exactly
the {\it true} covariance matrix.

\subsection{An uncorrelated model}
\label{sec:uncorr-model}

We are now going to show some first results for a simple uncorrelated model (i.e. diagonal $C$).
This is applied to the same data as Sec.~\ref{sec:traditional}, a pion correlator. Other ensembles and channels have been tried with similar results. We should stress that our data $y$ is labelled by configuration numbers, the likelihood being a product over configurations. This is different from the bootstrap analysis, where we first averaged the data inside each bootstrap sample and then worked in the space of averages, but bears some similarity with the fluctuations of the $\Gamma$ method. 

We choose our likelihood to be gaussian for simplicity. However,
here this is not a theoretical constraint: we crosschecked some of our results with a few arbitrary generalisations and even explain in Sec.~\ref{sec:BB} how to build a non-parametric inference.
This gaussian approximation is likely to work well as long as we keep working on correlators (no derived quantity), where the Central Limit Theorem applies to some extent, and with a large number of topological sectors. 
By no means it implies that the posterior distributions have to be gaussian\footnote{if all our fit parameters are precise enough for a Taylor expansion to apply,
it often means that we should have chosen a more complex model and would have signal on higher orders}.
In any case, any approximation we make here in the model-building can be evaluated a posteriori when comparing models according to Sec.~\ref{sec:IC}.

Having chosen some hyper-parameters $t_{min}$ and $N_{exp}$, we apply for the standard deviation a prior with a shape
\begin{equation}
  P(1/\sigma_t^2) = \Gamma(\alpha,\beta)
\end{equation}
with an arbitrary choice of $(\alpha,\beta)$ in Eq.~\ref{eq:gamma} inside the domain
\begin{equation}
  \alpha\ll n,\qquad\beta/\alpha=\frac{1}{n}\sum_{i=1}^n\left(y_i(t)-\bar y(t)\right)^2.
\end{equation}
This is flat for $\alpha=\beta=0$, and non-zero values are not actually needed for such a simple model.
The second piece we need to set is the likelihood
\begin{equation}
  P(y_{t>t_{min}}|v_j,E_j,\sigma_t) = {\cal N} (y_{t>t_{min}}|\sum_{j=1}^{N_{exp}} v_je^{-E_jt},\sigma_t),
\end{equation}
where arbitrary uninformative priors are set on 
$v_j$ and $E_j$, and the likelihood also contains the constraint that $E_j$ is ordered (this is implemented in PyMC through a reparametrisation).

\begin{figure}[ht]
\begin{center}
      \includegraphics[width=0.9\textwidth]{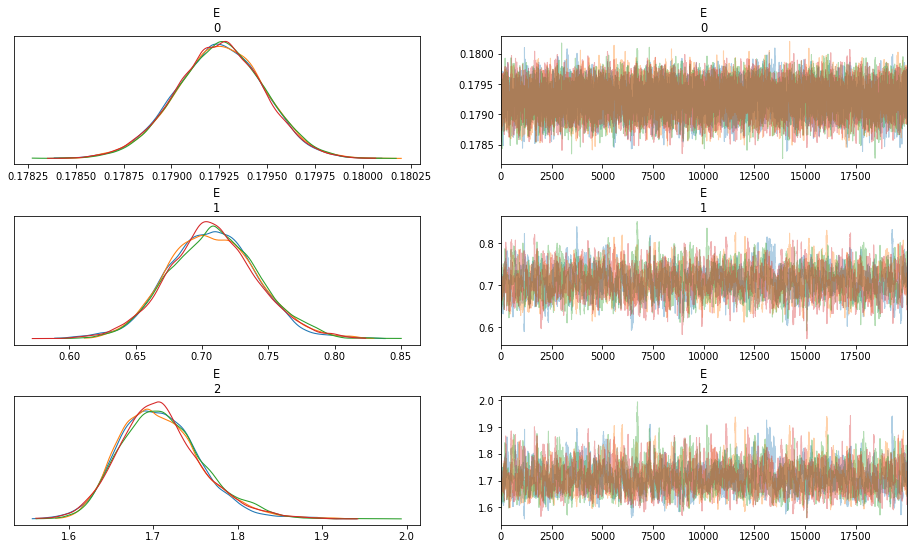}
\end{center}
\caption{We show the HMC trace of the uncorrelated model for the energy variables, one colour per chain. 
The left side represents the posterior distribution (value of the parameter on the x axis) for
each parameter while the right side represents the Monte-Carlo history (trajectory number on the x axis).
The small dispersion of the width of the posterior, between one chain and others, can be seen as the ``error on the error''.
}
\label{fig:flattraceNexp3}
\end{figure}

\begin{figure}[ht]
\begin{center}
      \includegraphics[width=0.8\textwidth]{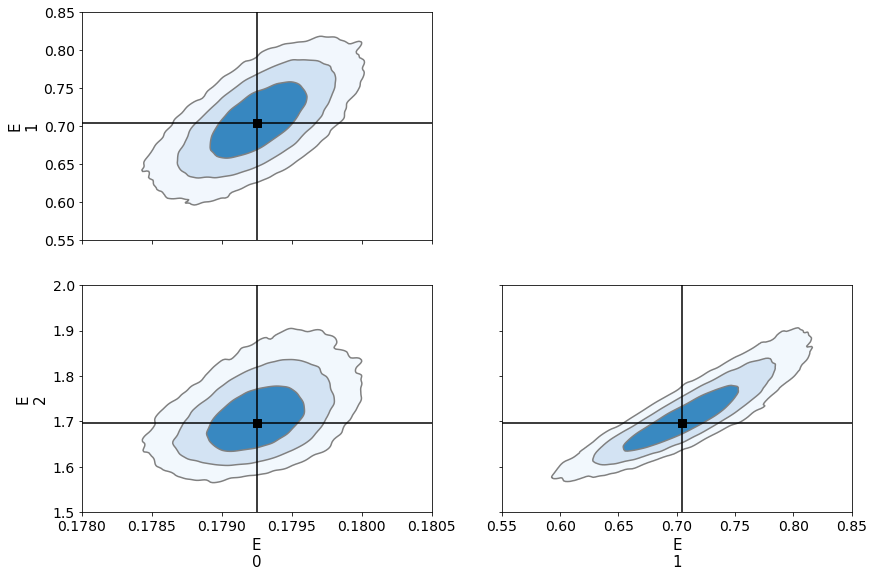}
\end{center}
\caption{We present the correlations between the densities of probabilities of each energy, 
in an uncorrelated model with unknown covariance and a very flat prior. The black square represents the mode of the posterior determined from the sampling, while direct minimisation suffers from the issues discussed
in Sec.~\ref{sec:traditional}. The 68\%, 95\% and 99.7\% contours are shown, by analogy with $1$, $2$ and
$3\sigma$.}
\label{fig:flatpairsNexp3}
\end{figure}

Once we have defined the model, we can choose for our analysis any level of complexity from a simple MAP to a full sampling
of the posterior. The results of the sampling are shown in Fig.~\ref{fig:flattraceNexp3}. In Fig.~\ref{fig:flatpairsNexp3} we compare
the results of these two methods with a more conventional $\chi^2$ fit.
The ground state is obtained at the per-mille precision while two extra
excited states appear to be well under control despite using extremely uninformative priors.
The trace is very useful to make sure our sampling was good enough, and provides a lot
of information compared to convergence metrics of MLE techniques. Here we notice how higher excited states are more difficult to sample, as the agreement between chains gets worse and autocorrelations get longer, but it is important that this is systematically
improvable by simply making the trace longer.

\subsection{A model with time correlations}

Similarly to Sec.~\ref{sec:uncorr-model}, we can define a correlated model with a multivariate distribution and the Wishart prior
of Sec.~\ref{sec:wishart}. This is implemented in PyMC through the Bartlett parametrisation but comes with an $O(p^2)$ increase
in the dimension of the sampling, with risks of slowdown and instabilities. We therefore propose a lightweight solution:
our choice of prior allows the $C$ variable to be marginalised analytically.
That is to say we want the total posterior probability of all combinations $(v,E,C)$ for a given value of $(v,E)$ but
{\it any} value of $C$:
\begin{equation}
P(v,E|y,M) = \int \frac{P(y|v,E,C,M)P(v,E|M)P(C|M)}{P(y|M)} dC.
\end{equation}
During the integration on $C$, all the parameters describing the mean are effectively {\it known} already, as arguments given
to the likelihood, and therefore the marginalised likelihood corresponds to the PPD of Eq.~\ref{eq:PPD-t}:
\begin{equation}
  P(y_{t>t_{min}}|v,E) = 
  t_{\nu+n-p+1}\left(y\mid\mu,\frac{\left(V^{-1}+\sum_{i=1}^n(y_i-\mu)(y_i-\mu)^T\right)^{-1}}{\nu+n-p+1}\right).
\end{equation}
where the known mean is substituted by its parametrisation
\begin{equation}
  \mu(t) = \sum_{j=1}^{N_{exp}} v_je^{-E_jt}.
\end{equation}

\begin{figure}[ht]
\begin{center}
      \includegraphics[width=0.9\textwidth]{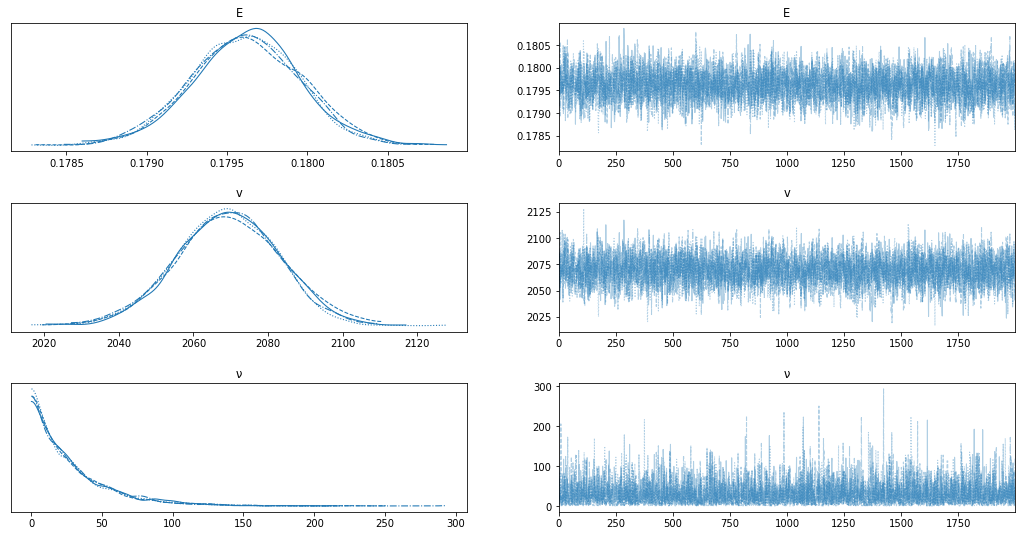}
\end{center}
\caption{We show the sampling of all parameters of the $N_{\rm exp}=1$ correlated model with a marginalised covariance matrix. 
Here different line styles represent each of the 4 chains. 
Typical values of $\nu$ are much lower than the number of
configurations, so that a very loose prior is selected. Note that the MAP could be determined from this sampling but, for the same model, our direct MAP computation failed. The
same model without marginalisation also tends to fail.}
\label{fig:margcorrtrace}
\end{figure}

\begin{figure}[ht]
\begin{center}
      \includegraphics[width=0.9\textwidth]{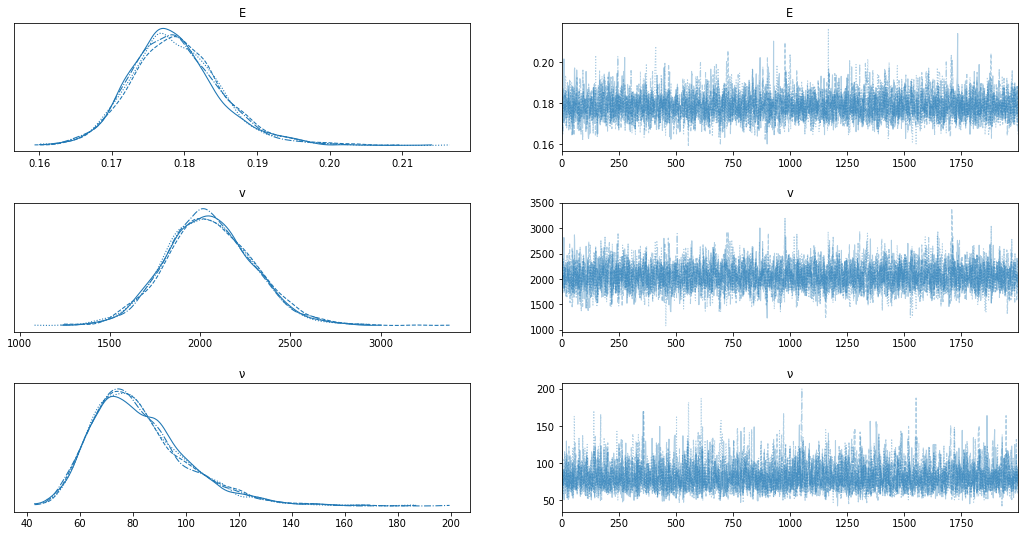}
\end{center}
\caption{Interestingly, the sampling is perfectly fine with using only $9$ configurations. In this case we can even compute the MAP. By contrast, it is known that with such small statistics
at least half the eigenvalues of the empirical covariance matrix of the data are algebraically zero, so that a $\chi^2$ fit would not be possible.}
\label{fig:margcorrtrace9cfg}
\end{figure}

In Fig.~\ref{fig:margcorrtrace} we show some results of the sampling for the marginalised model, and in Fig.~\ref{fig:margcorrtrace9cfg} we show the same thing for extremely low statistics.

Unfortunately the model without marginalisation could not be sampled with decent ergodicity and
auto-correlations\footnote{The Wishart and Wishart-Bartlett
implementations in PyMC are known to be inefficient and the documentation recommends using a LKJ prior instead.
Later versions or other software might be more efficient.}, so we cannot check explicitly our intuition that using the full likelihood (including the normalisation factor that a $\chi^2$ fit
does not take into account) allows our inference to pick covariance matrices with better condition numbers and more meaningful $\chi^2$.

\subsection{A model with auto-correlations}

If we want our bayesian analysis to supersede both bootstrapping and the $\Gamma$ method, we need to include in our
model some support of auto-correlated data. This class of problems is known as {\it time series analysis}.

Binning is always a possibility, and will be discussed in Sec.~\ref{sec:binning}, but in this section we propose a more direct solution.

Very little is known theoretically about the spectrum of the HMC as a Markov process, which makes it difficult to come
up with a valid modelisation. However, it is reasonable to require our model to have some exponential decay of
autocorrelation at long distances, and to include a few modes to distinguish $\tau_{int}$ from $\tau_{exp}$. This is
what the auto-regressive AR(r) models provide: given some normally distributed innovation $\xi={\cal N}(0,\sigma)$
we define
\begin{equation}
y(\tau) = \rho_0 + \sum_{i=1}^r \rho_i y(\tau-i) + \xi,
\end{equation}
where $\tau$ stands for the HMC time, as opposed to the Euclidian time $t$.

The covariance of $y$ in this model could be written as a function of the $\rho$ elements, and cutting this recurrence to a given
order is the analogue of choosing a window for the $\Gamma$ method.
This would involve computing the inverse of a band-diagonal matrix, so even when $r$ is small there are 
long-range correlations in $\tau$.

\begin{figure}[ht]
\begin{center}
      \includegraphics[width=0.9\textwidth]{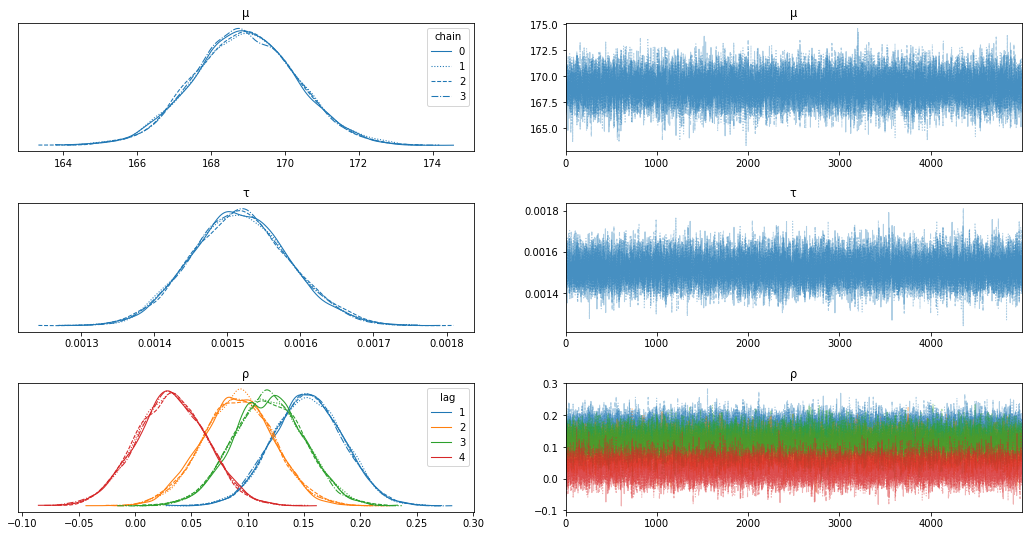}
\end{center}
\caption{This $AR(4)$ model provides a result $\mu=168.9(1.4)$ for the expected value of the correlator at this time slice. A naive estimate neglecting auto-correlations would underestimate the error as
$\mu=168.9(8)$.
The signal remains stable when adding higher orders, but those would be largely compatible with zero.}
\label{fig:ar4trace}
\end{figure}

\begin{figure}[ht]
\begin{center}
      \includegraphics[width=0.7\textwidth]{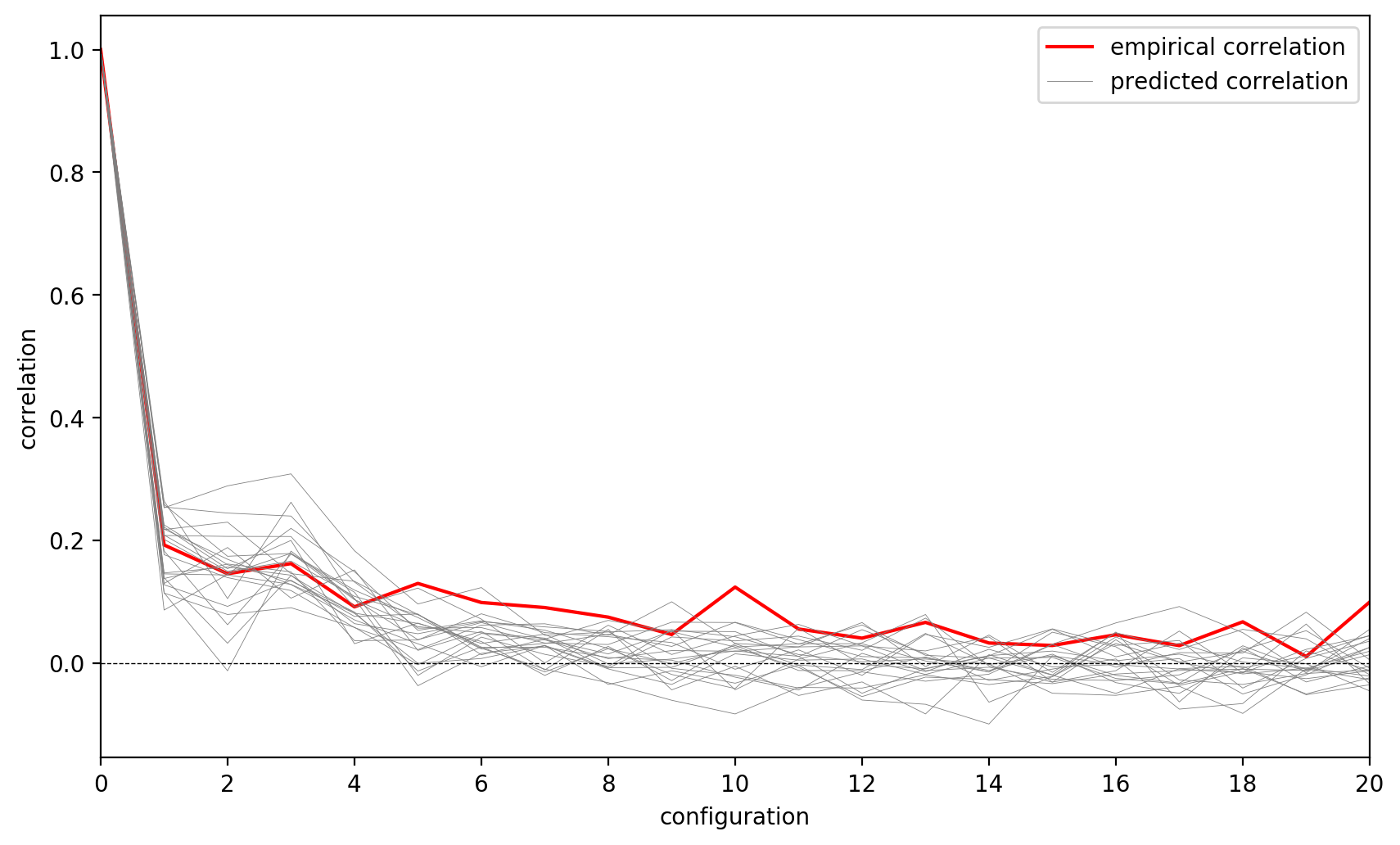}
\end{center}
\caption{As we use the sampled parameters to generate predictions from this $AR(4)$ model, we can check that the auto-correlation of these predictions look very similar to the auto-correlations of the data.
Together with the values of $\rho$ this gives a confirmation that the order is high enough.}
\label{fig:ar4corr}
\end{figure}

Fig.~\ref{fig:ar4trace} shows the result for an $AR(4)$ model to describe a correlator on a single time slice.
In Fig.~\ref{fig:ar4corr} we compare the auto-correlation of the data and the PPD.

\subsection{Perspectives for a full model}

As we want to take into account both the correlation in HMC time and in Euclidian time, we face the curse
of dimensionality. Vector auto-regressive (VAR) models are a sensible description, where $\xi$ is multivariate
and each $\rho$ parameter
is a $p\times p$ variable, but this comes with $O(p^2r)$ parameters and is unlikely to be realistic.

\begin{figure}[ht]
\begin{center}
      \includegraphics[width=0.9\textwidth]{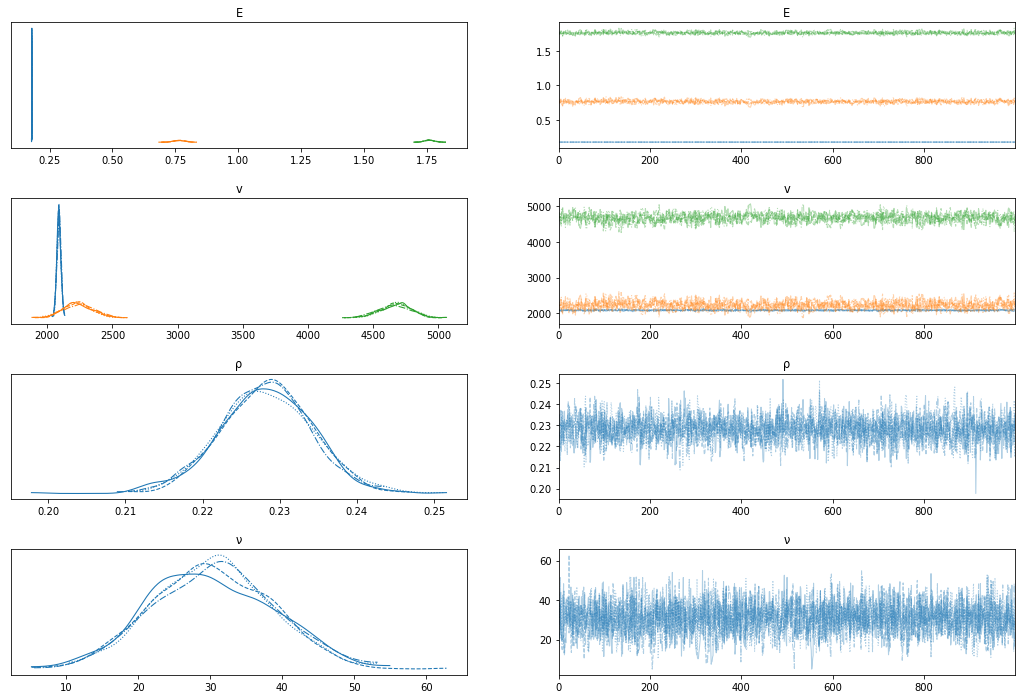}
\end{center}
\caption{This model is similar to Fig.~\ref{fig:margcorrtrace} except that it is based on an AR(1) model taking into account some auto-correlations.
The error bar on $E_0$ is therefore about $20\ \%$ more conservative, which is consistent with the lag parameter $\rho$ being clearly non-zero.}
\label{fig:VARtrace}
\end{figure}

Therefore some level of approximation is needed, which is nothing but an underfitting-overfitting trade-off and
should be evaluated according to the methods of Sec.~\ref{sec:IC}. The most relevant model depends on each data set
so we will not discuss this in much details here. However, we show some results in Fig.~\ref{fig:VARtrace} for a very simple vector AR(1) model
whose matrix is assumed to be constant.

Note that a similar issue exists for the $\Gamma$ method, in which computing the covariance with auto-correlations taken
into account (differently for each matrix element) is difficult and led to the common use of uncorrelated fits.

\section{Information criteria and model averaging}
\label{sec:IC}
\subsection{From Akaike to Watanabe-Akaike}

The Akaike criterion\cite{akaike1976canonical}
\begin{equation}
AIC = - 2 \log {\cal L}(a_{MLE}) + 2k 
\end{equation}
became popular in the lattice community\cite{Jay:2020jkz}, since it can be related to a correlated $\chi^2$.
This is turned into a weight
\begin{equation}
w(M) = \frac{e^{-AIC/2}}{\sum_{M'} e^{-AIC(M')/2}}
\end{equation}
which asymptotically describes the marginal probability of model $M$. The value of a parameter $a_c$
common to all models would be distributed with
\begin{equation}
P(a_c) = \sum_M w(m) P(a_c|M).
\end{equation}

This is sometimes called pseudo-bayesian model average, since it uses an approximation around the MLE while a genuinely
bayesian model average should use some marginal probabilities as weights.

The Watanabe-Akaike\cite{WATANABE_2010} or widely applicable information criterion (WAIC) is a bayesian generalisation
 that takes into account the full posterior as
\begin{equation}
WAIC(M) = - \sum_{i=1}^n \log P(y'_i|y,M) + k_{WAIC} ,
\label{eq:WAIC}
\end{equation}
where the effective number of parameters\footnote{This is the $p_{WAIC2}$ implemented in PyMC, which is slightly different from $p_{WAIC1}$ originally proposed by Watanabe}
\begin{equation}
k_{WAIC} = \sum_{i=1}^n \left\{ \int \left(\log P(y'_i|a,M)\right)^2 P(a|y,M) da
-\left(  \int \log P(y'_i|a,M) P(a|y,M) da \right)^2 \right\}
\label{eq:kWAIC}
\end{equation}
goes to $k$ in the domain of validity of the AIC\footnote{Note that conventions for AIC and WAIC differ by a factor $2$ in
\cite{WATANABE_2010} and $-2$ in \cite{Vehtari_2016}}.

This criterion can be applied to almost any model, with very weak assumptions
about how close this is to the {\it true} model, unlike the AIC which requires to be in its neighbourhood. This means
for instance we can compare the WAIC value between a correlated and an uncorrelated fit.

While, by using those information criteria, we seem to have completely abandoned the concept of {\it goodness-of-fit}, it is worth noting that the traditional condition $\chi^2\leq p-k$
is strictly equivalent to an inequality $AIC_{\chi^2}\le AIC_{trivial}$ describing whether the $\chi^2$ model of Sec.~\ref{sec:reinterpret} makes better predictions than
the trivial model of Sec.~\ref{sec:trivial}.

\subsection{Cross-validation and the elpd}

The WAIC happens to be very close to another estimator, built from the leave-one-out cross-validation (LOO)\cite{Vehtari_2016}.
Those are actually two approximations of the expected log predictive distribution (elpd)
\begin{equation}
elpd = \sum_{i=1}^n\int P(y'_i|M^*)\log P(y'_i|y,M) dy'_i
\end{equation}
where $M^*$ stands for the {\it true} model which generated the data.
In PyMC the LOO is obtained
from importance sampling (aka reweighting) of the posterior, so that one does not need to regenerate a new HMC
chain for each separation between the training and the test data.

The elpd itself is related to the cross-entropy so that the maximisation of the elpd is also a
minimisation of the Kullback-Leibler divergence from $M$ to $M^*$
\begin{equation}
D_{KL}(M^*||M) = \sum_{i=1}^n \int P(y'_i|M^*)\log\frac{P(y'_i|y,M^*)}{P(y'_i|y,M)} dy'_i.
\end{equation}

Finally, let us consider the case where the true model is ${\it known}$ to be a multivariate normal distribution
with a {\it known} covariance (e.g. because it is synthetic data) and our model is an uncorrelated fit: 
then the elpd is very similar to the the expected $\chi^2$
of \cite{Bruno:2022mfy}. The difference is that the elpd averages over the full posterior (predictive) instead of only looking at the
divergence from the MLE.

\subsection{Categorical variables and mixture models}

As bayesian models can in principle include any level of complexity, an interesting equivalent of BMA is writing a single model which is general enough to include our full set of models.
One way to do this is by adding some categorical variables controlling which model is used.
For instance
\begin{eqnarray}
  P(t_{min}) &=& {\mathbbm 1}_{t_0<t_{min}<t_1}\\
  P(y_{t<t_{min}}|\mu_t,\sigma_t) &=& {\cal N} (\mu_t,\sigma_t)\\
  P(y_{t\ge t_{min}}|v,E,\sigma_t) &=& {\cal N} (\sum_{j=1}^{N_{exp}} v_je^{-E_jt},\sigma_t).
\end{eqnarray}

The posterior probability $P(t_{min}\mid y)$ obtained in the trace can simply be interpreted as the BMA weight.

Note however that the explicit use of categorical variables often destabilises the HMC, so marginalising them (which is trivial for a finite set) is usually a good idea. It also ensures that all values of $t_{min}$
contribute to the sampling at each step, even the rarest.
The model is then represented as a mixture
\begin{equation}
P(y|\mu,v,E,\sigma) \propto \sum_{t_{min}} \prod_{i=1}^n\left\{ \left[
  \prod_{t<t_{min}} P(y_{it}|\mu_t,\sigma_t) \right]\left[
  \prod_{t\ge t_{min}} P(y_{it}|v,E,\sigma_t)
\right]\right\}.
\end{equation}
We also choose to add in our hierarchical model an extra layer with an arbitrary (e.g. $\alpha=1$) Dirichlet hyperprior
\begin{eqnarray}
  P(w) &=& {\cal D}(\alpha,\cdots,\alpha)\\
  P(t_{min}) &=& w_{t_{min}} .
\end{eqnarray}
Once $t_{min}$ is marginalised, the continuous variables $w$ can still be traced.

If, instead of marginalising $t_{min}$ or $w$, one first marginalises all the other parameters $\mu_t,v_j,E_j,\sigma_t$, the equivalence with BMA 
becomes manifest. The only difference is
the explicit presence of a $w$ prior instead of an implicit flat prior.

\begin{figure}[ht]
\begin{center}
      \includegraphics[width=0.65\textwidth]{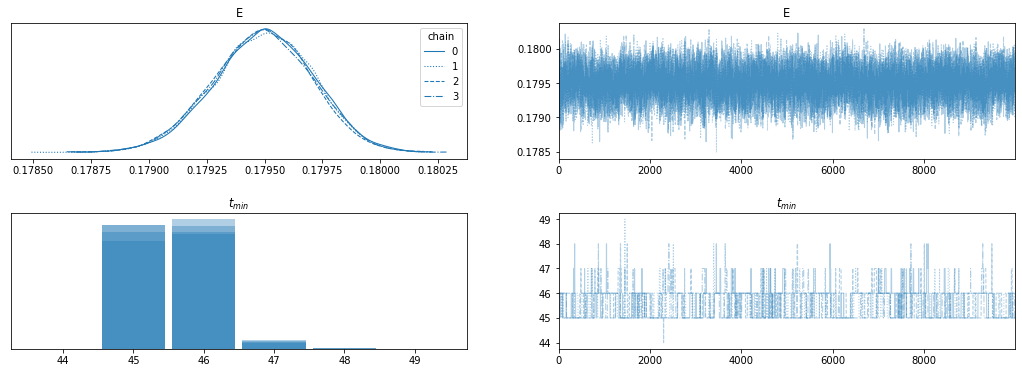}
      \hfill
      \includegraphics[width=0.34\textwidth]{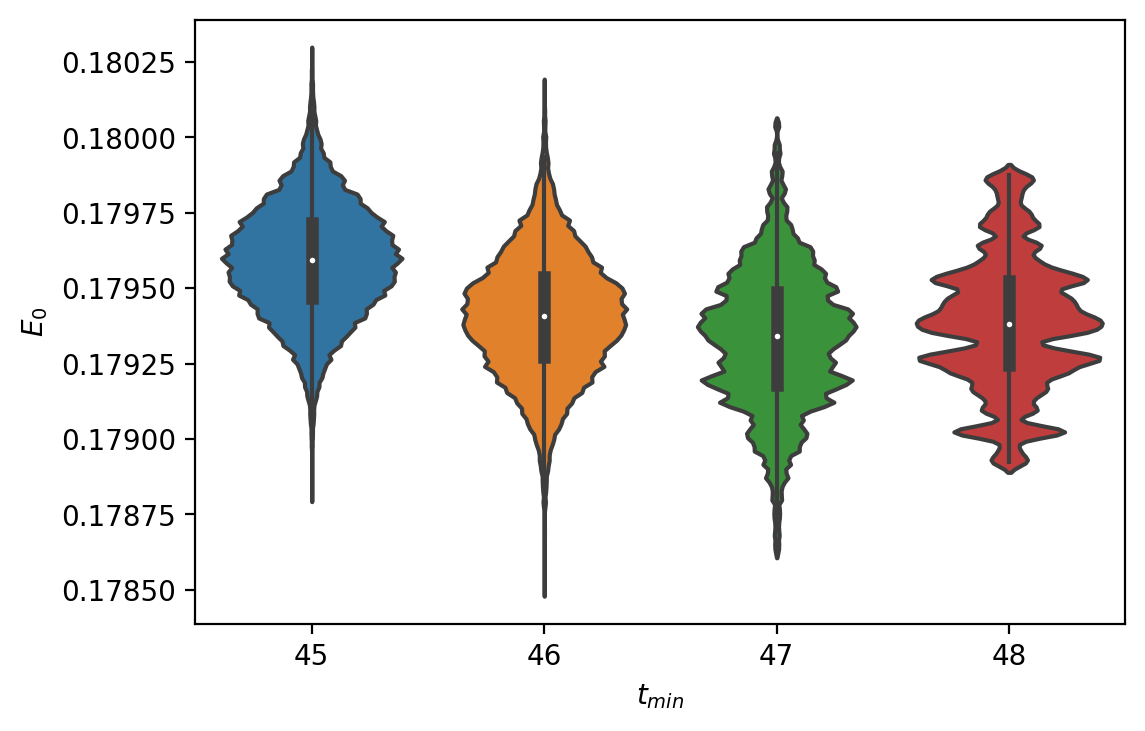}
\end{center}
\caption{The non marginalised model is more difficult to sample but gives a direct interpretation, where the posterior probability of $t_{min}$ corresponds to the BMA weight.
As shown in the violins on the right side, one can also look a posteriori at the sub-samples separately for each $t_{min}$. 
We see that (unlike when model averaging a posteriori individual samplings or fits) the sampler spends more time to describe precisely
models which actually matter, which could be a good or a bad thing.
}
\label{fig:tracenonmargmixture}
\end{figure}

\begin{figure}[ht]
\begin{center}
      \includegraphics[width=0.9\textwidth]{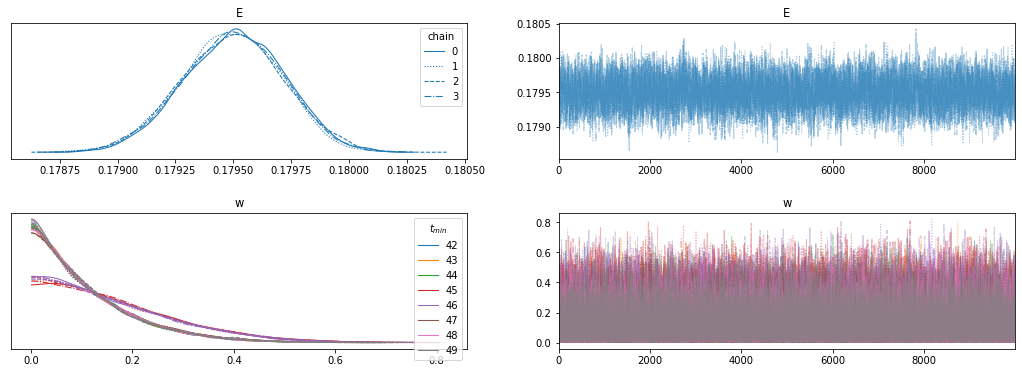}
\end{center}
\caption{This marginalised model leaves us with continuous variables $w$ which still provide information about $t_{min}$ but are more delicate to interpret. In particular we see that $w$ gives slightly more weight to
$t_{min}=45$ and $46$.
The results are very similar to Fig.~\ref{fig:tracenonmargmixture}, both in terms of $E$ and $w$.
}
\label{fig:tracemargmixture}
\end{figure}

Some results are shown in Fig.~\ref{fig:tracenonmargmixture} for the non-marginalised version and Fig.~\ref{fig:tracemargmixture} for the marginalised version, which should give the same
results up to practical issues.
In both cases the models combined here are derived from the uncorrelated model of Fig.~\ref{fig:flattraceNexp3}, where $t_{min}$ cuts have been introduced.

\subsection{A comment on Bayes factors}

In principle, one can compute the marginal probability of a model directly, integrating the denominator of
Eq.~\ref{eq:bayes}. There would then be no need for an information criterion, although it has been shown to be
asymptotically equivalent to the BIC.
The relative importance of two models is given by the Bayes factor
\begin{equation}
K(M_1,M_2) = \frac{P(y|M_1)}{P(y|M_2)}
\end{equation}

\begin{figure}[ht]
\begin{center}
      \includegraphics[width=0.6\textwidth]{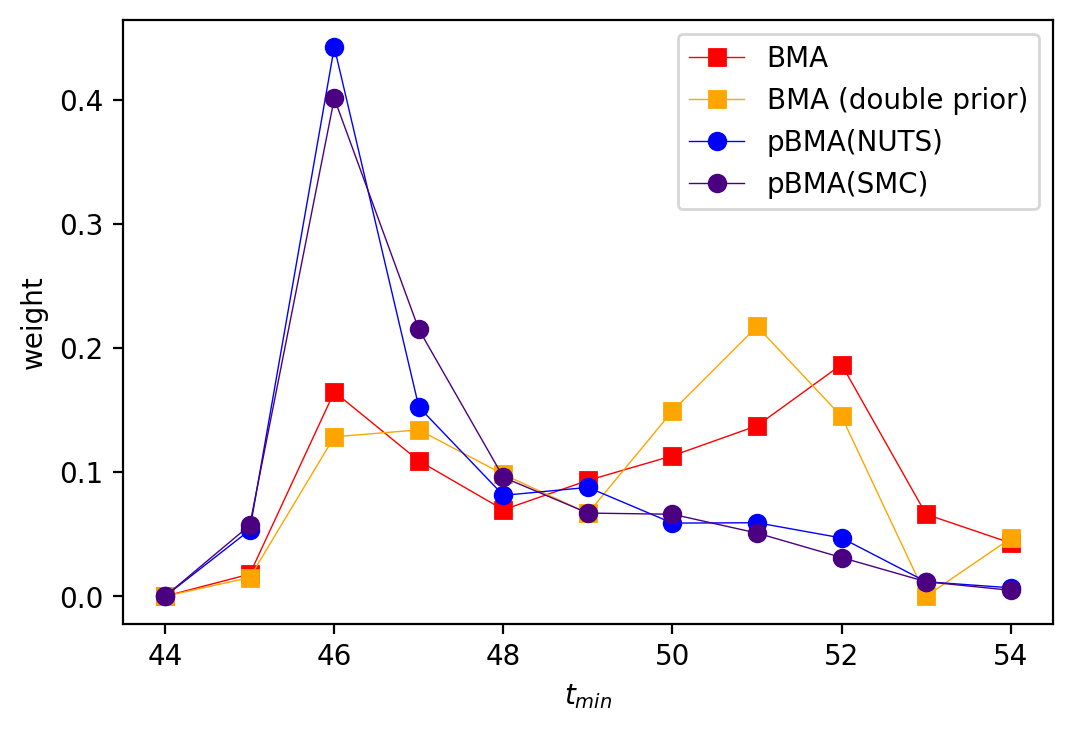}
\end{center}
\caption{For the same problem as Fig.~\ref{fig:tracemargmixture} we now look at a direct computation of the BMA weights from the marginal likelihoods computed
with Sequential Monte Carlo (SMC). We also show the comparison with the pBMA weights (from the WAIC). Those are different objects since in theory the marginal likelihood
asymptotically gives the BIC while the WAIC asymptotically gives the AIC. We also compare two choices of prior.
}
\label{fig:BMAvspBMA}
\end{figure}

This can be computationally intensive and less stable, but is doable within PyMC as shown in Fig.~\ref{fig:BMAvspBMA}.
Instead of the usual variant of the HMC used by PyMC, called NUTS, this uses a more complicated procedure called
Sequential Monte-Carlo (SMC) to solve the problem in several steps.
One should note that marginal probabilities, which are related to the prior predictive distribution, can in principle have a strong sensitivity
to the choice of prior. We checked that it is not the case in our specific example.

\section{Misspecified model}
\label{sec:misspecified}

\subsection{Asymptotic knowledge of covariance}

We discussed in Sec.~\ref{sec:reinterpret} how bad a $\chi^2$ fit is when considered as a bayesian model, but showed in
Sec.~\ref{sec:wishart} how it is asymptotically similar to a well-defined model if the average is modelled close to the truth.
We then introduced in Sec.~\ref{sec:IC} an information criterion built on some very weak assumptions. This WAIC can even be applied
to misspecified models, where the AIC can not. This allows for an empirical check of the convergence of the WAIC towards the AIC, as well as comparing $\chi^2$ fits and the full bayesian model on the basis on the WAIC.

Determining the AIC can be difficult for complicated models, where minimising is much more unstable than
sampling, however comparing $k_{WAIC}$ to $k$ is usual easier. In our experience, $k_{WAIC}$ appears to
always stay very close to $k$ for $\chi^2$ fits, even for low statistics.
This is probably related to the discussion of Sec.~\ref{sec:WAIC-cst}.
So, by this criterion, the asymptotically small misspecification of $\chi^2$ models does not seem to
have a large impact.
However, when comparing WAICs versus the full bayesian model, the $\chi^2$ model becomes strongly
excluded in the limit of large statistics.

\begin{figure}[ht]
\begin{center}
      \includegraphics[width=0.4\textwidth]{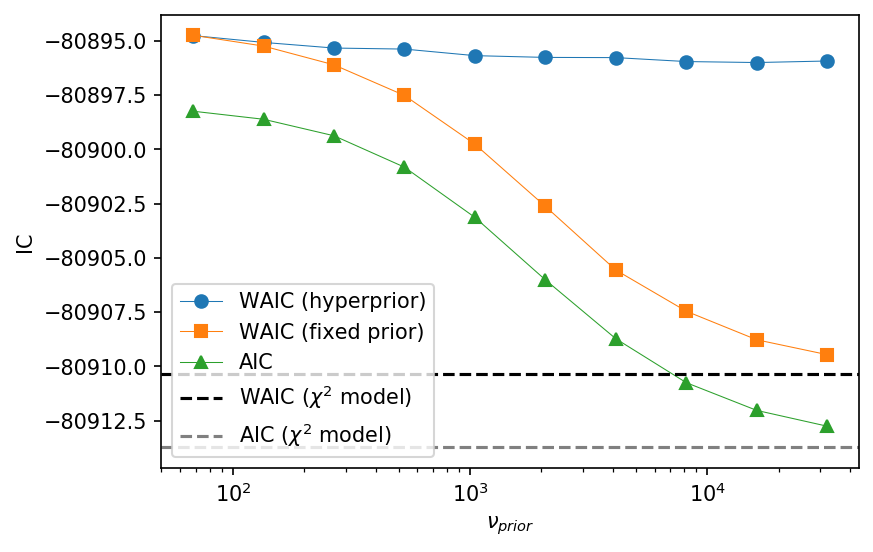}
      \hfill
      \includegraphics[width=0.4\textwidth]{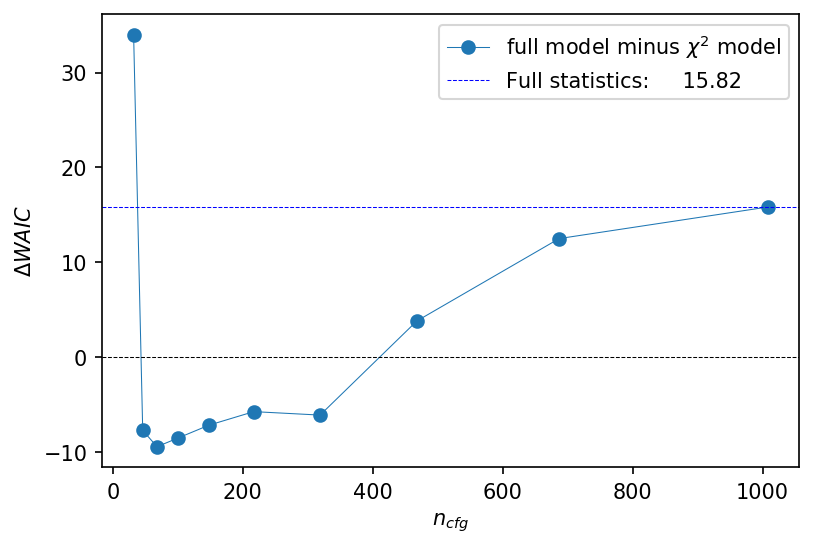}
\end{center}
\caption{
We show the information criteria for our bayesian model, using the prior of Sec.~\ref{sec:wishart},
as a function of the strength of this prior (left, full statistics) or the number of configurations
included in the analysis (right, hyperprior with $\nu_{\rm prior}=\sqrt{n_{\rm cfg}}$).
The scale matrix in the prior is set to the empirical covariance. The $\chi^2$
model ($\nu$ fixed to $\infty$) is shown as a comparison.
For large enough statistics, looser priors are favoured by the WAIC (and selected by the hyperprior), so
that it strongly excludes the $\chi^2$ model where $C$ is totally fixed a priori.}
\label{fig:nuprior}
\end{figure}

\begin{figure}[ht]
\begin{center}
      \includegraphics[width=0.4\textwidth]{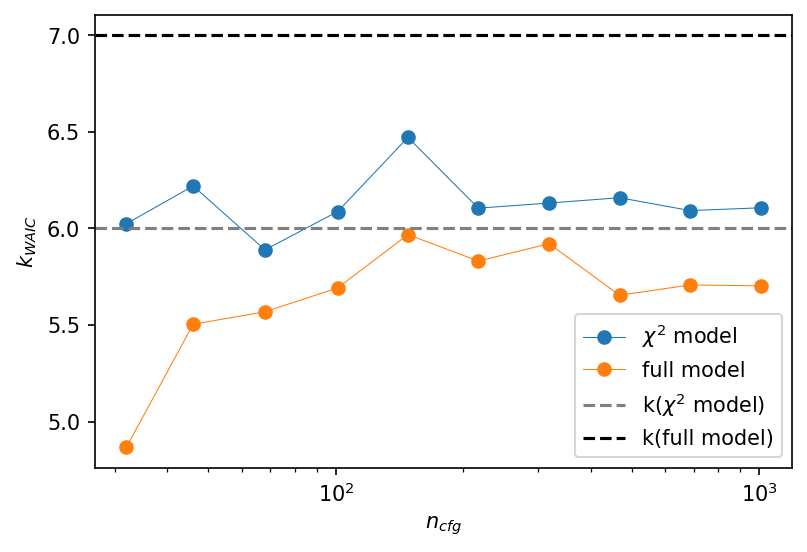}
      \hfill
      \includegraphics[width=0.4\textwidth]{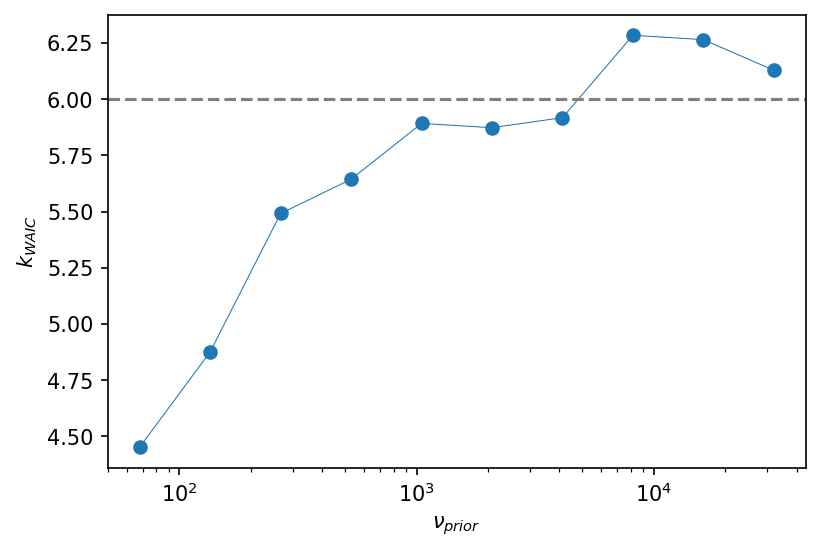}
\end{center}
\caption{We show the effective number of degrees of freedom of the WAIC, as a function of the number of
configurations included in the analysis (left) or the strength of the prior (right, $100$ configurations).
For looser $\nu$ priors and low statistics, the effective number of parameters tends to drop down.
This usually means that for some parameters we have no signal and nothing new is learnt beyond the priors.}
\label{fig:pnuprior}
\end{figure}

In Fig.~\ref{fig:nuprior} (resp. Fig.~\ref{fig:pnuprior}) we show how our information criteria
(resp. the effective number of parameters)
vary when we change the strength of a prior acting on the covariance.
A similar study is done when the number of configurations changes.
In one model we define the hyperprior $P(\nu)=\exp(-\nu/\nu_{\rm prior})$ where $\nu$ is the Wishart parameter of the model of Sec.~\ref{sec:uncorr-model}.
In another model we fix $\nu$ exactly to $\nu_{\rm prior}$. 
The scale parameter $V$ is chosen so that the Wishart prior is centered around the empirical covariance of the data.
For $\nu_{\rm prior}$ small the prior gets flat, and therefore this extra freedom allows the model to parametrise the true distribution.
On the other hand for $\nu_{\rm prior}\to\infty$ the model tends toward the $\chi^2$ model of Sec.~\ref{sec:reinterpret}, which is misspecified.

Note that it is not immediately clear what are the parameters for a hierarchical model such as this one.
For the AIC we only count the degrees of freedom
which are minimised, excluding those which have been integrated out.

\subsection{Information criteria and biased covariance}

Let us consider the case where we know the actual covariance matrix $C$ but do not want to use it in our model, for instance because of stability issues.
We decide to fit our data to a gaussian model (a $\chi^2$ fit) with an arbitrary weight $W$. The model is then misspecified in $W$, but at the same
time we assume that the parametrisation of its mean is still correct, and its {\it true} value is asymptotically reached.

This typically happens when performing an uncorrelated fit (diagonal $W$) on correlated data.

In this case the AIC does not apply. The WAIC, however, does.
In this section we will look at the Takeuchi Information Criterion\cite{Takeuchi76} (TIC), another criterion with an intermediate level of applicability,
which is asymptotically equivalent to the WAIC but can be used on MLE just like the AIC.
Another advantage is that deriving a close formula is easier for the TIC. We start from the formula presented in English in
\cite{WATANABE_2010} and write the effective number of parameters
\begin{eqnarray}
k_{TIC} &=& \Tr\left(I(a_{MLE})J^{-1}(a_{MLE})\right)\\
I(a) &=& \int \nabla\log P(y\mid a,M)\nabla\log P(y\mid a,M) P(y\mid M^*) dy\\
J(a) &=& -\int \nabla^2\log P(y\mid a,M) P(y\mid M^*) dy .
\end{eqnarray}

The assumptions written at the start of this section translate into
\begin{eqnarray}
P(y\mid a,M) &=& \frac{\det(W^\dag W)^{1/2}}{(2\pi)^{p/2}}\exp{\left(-\frac{1}{2}||W(y-f(a))||^2\right)}\\
P(y\mid M^*) &=& \frac{\det(C)^{-1/2}}{(2\pi)^{p/2}}\exp{\left(-\frac{1}{2}(y-\mu)^TC^{-1}(y-\mu)\right)}.
\label{eq:TIClikelihoods}
\end{eqnarray}

Gaussian integrals can then be worked out analytically to find that if $|f(a_{MLE})-\mu|=O(1/\sqrt{n})$ can be neglected then
\begin{equation}
k_{TIC} \simeq \Tr\left[(G^\dag C_W G)(G^\dag G)^{-1}\right] = \Tr\left[ {\cal P} C_W \right],
\label{eq:kTIC}
\end{equation}
where $C_W = WCW^\dag$, $F_{i\alpha}=\partial f_i/\partial a_\alpha$ ($i\le p, \alpha\le k$) is the $n\times k$ Jacobian matrix of the fitting function, $G=WF$ and ${\cal P}=G(G^\dag G)^{-1}G^\dag$ is a projector.
Those are the same quantities which appear in the expected chi square\cite{Bruno:2022mfy}, and the TIC can also be written, up to a constant which does not depend on $f$, as
\begin{equation}
TIC = \chi^2_{MLE} - 2 E(\chi^2|M^*).
\end{equation}

If $W=C^{-1/2}$ then our model is well-specified and we recover the AIC as $k_{TIC}=k$.
If $W_{ij}=(C_{ii})^{-1/2}\delta_{ij}$ and the data is strongly correlated, then $k_{TIC}\to p$, which compensates the fact that uncorrelated models tend to underestimate the $\chi^2$ entering
the first term of the information criterion. Eq.~(\ref{eq:kTIC}) has been cross-checked in terms of
its good agreement (up to a few percent) with the full numerical computation of $k_{WAIC}$.

A third interesting case is $W=\sqrt{\beta} C^{-1/2}$, where $\beta$ can be interpreted as an inverse temperature.
The temperature does not change the position of the MLE, but it does change the value of the likelihood at its maximum.
It also gives $k_{TIC}=\beta k$.
At zero temperature, where the system is frozen at its MLE, the model is penalised by an increase of $k_{TIC}$: statistical
fluctuations are amplified in the likelihood and this leads to overfitting.
At high temperature the posterior distribution learns nothing about the fit parameters, so the log likelihood is diluted
over many values of the parameters and this results in bad values of the TIC (or WAIC) as well.
In Tab.~\ref{tab:waic-vs-aic} we show a comparison between the AIC and the WAIC on an uncorrelated model with various values of $\beta$ in $W_{ij}=\sqrt{\beta} (C_{ii})^{-1/2}\delta_{ij}$.

The same formula applies to $k_{WAIC}$ in the large $n$ limit for constant fits (and probably any fitting function $f$ linear in its parameters) with flat priors, as shown in Sec.~\ref{sec:WAIC-cst}.

\begin{table}
\begin{center}
\begin{tabular}{ccccc}
\hline
model & $\Delta_{WAIC}$ & $k_{WAIC}$ & AIC-WAIC & $E$ \\
\hline
$\beta=1$ & & 16.79 & 8.2 & 0.17938(28)\\
$\beta=0.5$ & 1651 & 8.25 & 3.9 & 0.17936(39)\\
$\beta=2$ & 2642 & 32.62 & 16.2 & 0.17938(20)\\
$\beta=0.1259$ & 10269 & 2.04 & 0.81 & 0.17937(78)\\
\hline
\end{tabular}
\end{center}
\caption{For an uncorrelated $\chi^2$ model with a single exponential starting at $t_{src}+13$ we compare several models. $\beta=1$ is
the standard bayesian inference, which is strongly preferred in terms of WAIC, despite a large $k_{WAIC}$ coming from strongly correlated
data. The low-temperature case $\beta=2$ compresses the result closer to the MLE, with smaller error bars
resulting in a model
with an actually lower predictive power. The case $\beta=0.1259$ is tuned to give $k_{TIC}=k$ (so that
$\beta$ compensates our dropping the off-diagonal elements of $C$), and this does not result in a good model
neither, as too much conservatism is penalised. 
WAIC and AIC do not agree here since the models are misspecified in $C$.
The very small differences in the central value of $E$ come mostly from systematically-improvable Monte-Carlo errors.}
\label{tab:waic-vs-aic}
\end{table}

\section{Conclusion}

We presented some fully bayesian framework that we put in practice on a pion correlator to perform a complete analysis, without any extra ingredient or layer.
In this proof-of-concept we chose to stick to a relatively simple problem. However it turned out to outperform classical methods in the sense that we could
reliably fit three states with high precision, without cutting any early-time data, or perform correlated fits with 9 configurations.
The superiority of our new method, 
as well as benefits from its flexibility,
is likely to be more obvious for more complicated problems which we plan to study in the future.

However, this method is meant to make the modelling assumptions more visible and assessable rather than to automatically solve all the challenges we face. Therefore, one should on a case-by-case
build and evaluate models to correctly describe our data. We present a few possibilities to tackle some of the issues commonly arising in Lattice QCD data, in particular
Euclidian time and Monte-Carlo time correlations. More work is necessary to identify good models for various situations where several of these complications happen at the same time.

We discussed some of the tools at our disposal for this evaluation of models, including both bayesian and MLE-based information criteria, and presented a few ways to
perform model averaging.

As the code we wrote is being made public, we invite the reader to try our models on their own data or build more ambitious models.
The same models could also, once we start from a well-defined bayesian problem, be used on several intermediate level of approximation.
Many are already implemented in the PyMC package we used, and it can make sense to use one or the other depending on practical constraints and the complexity of the problem.
Having all those in a single toolbox can certainly be convenient: hierarchical models, gaussian likelihoods, variational inference, normal approximation around the MAP, \dots

\appendix

\section{WAIC of a constant fit}
\label{sec:WAIC-cst}

A constant fit is a particular case of Eq.~\ref{eq:TIClikelihoods} with $f(a)=aE$, where $E=(1,\cdots,1)$.
The effective number of parameters $k_{WAIC}$ is then given by the variance of
\begin{equation}
\log\left(P(y_i | a)\right) = -\frac{1}{2} || W(y_i-aE) ||^2
\end{equation}
according to the posterior distribution of $a$.

With a flat prior, this distribution can be computed analytically as
\begin{eqnarray}
P(a|y) &\propto& P(y|a) \propto \prod_i \exp\left(-\frac{1}{2} || W(y_i-aE) ||^2\right)\\
  &\propto& \exp\left(-\frac{|WE||^2}{2n}\left(a-\frac{\langle W\bar{y}\mid WE\rangle}{||WE||^2}\right)^2\right),
\end{eqnarray}
where $\bar{y}$ stands for the average of all vectors $y_i$.

We then expand the log-likelihood around the mean of this posterior, getting rid of the leading terms which have no variance in $a$
as well as of the asymptotically negligible $(a-\bar{a})^2$ terms:
\begin{equation}
\log\left(P(y_i | a)\right) \simeq {\mathrm{cst}} + \langle Wy_i-\frac{\langle W\bar{y}\mid WE\rangle}{||WE||^2}WE\mid
                                 aWE-\frac{\langle W\bar{y}\mid WE\rangle}{||WE||^2}E \rangle.
\end{equation}

Since we know the variance of $a$ this now boils down to
\begin{equation}
V\left(\log\left[P(y_i | a)\right]\mid M,y\right) \simeq \frac{\langle W(y_i-\bar{y})\mid WE\rangle^2}{n||WE||^2},
\end{equation}
and finally once summed on $i$, if in this section $C$ is the empirical covariance of the data,
\begin{equation}
k_{WAIC} \simeq \frac{\langle WE\mid C_W\mid WE\rangle}{\langle WE\mid WE\rangle},
\end{equation}
which is almost exactly Eq.~\ref{eq:kTIC}.

\section{The bayesian bootstrap}
\label{sec:BB}

An apparent limitation of our method is that we need to provide a parametric model of the noise.
In reality, not only this model can be arbitrary complicated but non-parametric models exist as well.
We are in particular going to present the bayesian bootstrap (BB)\cite{Rubin81}, not because we are advocating for its use but because of its similarity with the
classical bootstrap\cite{efron1992bootstrap} commonly used in the LQFT community.

Let us imagine we are only interested in a single number, like a plaquette or a correlator at a fixed time. We
can write a simple model as follows:
\begin{eqnarray}
P(w) &=& {\cal D}(1,1,\cdots,1)\\
P(y) &=& \sum_i w_i\delta(y-y_i),
\end{eqnarray}
where ${\cal D}$ is the Dirichlet distribution used as a prior on $w$.
This formula means that our likelihood gives a probability $w_i$ to observe a value similar to what we have already
observed in the data and zero for any other value. This is similar to having a probability $1/n$ to pick $y_i$
to build a classical bootstrap sample, except that more freedom is given to $w$ and smoother distributions are obtained.

Note that this model can be seen as the marginalised version of a data-independent model giving weight to {\it all} values
of $y$, integrating out trivially all the unobserved values.

The difficulty comes when trying to replace our usual bootstrapped fit by a model putting together the
bayesian bootstrap with a parametric model giving explicit access to physically relevant quantities.
Indeed, there is some contradiction in the fact that on the one hand the BB refuses to make any generalisation and to
include in the likelihood any value other that those that we already observed, and on the other hand the fact that
we want a parametrisation which allows for physical predictions on new systems.
This is not a new problem, since it was already somewhat unnatural that the result of the classical
bootstrap is usually fed to a least square based on a strong gaussian assumption.
We believe this can be solved by choosing a less extreme model with some flexibility, halfway between one single gaussian and an infinity of Dirac distributions, but leave that to future work.

\section{Global fits}
Fits are often chained in a LQFT analysis, for instance by performing first a correlator fit and then a continuum limit on the
results of those fits, combined with scale-setting inputs from yet another fit. 
Methods have been developed to properly propagate the correlations between various observables through this chain,
when each observable is fitted independently and then added with the others in a global fit. In a bootstrap analysis for instance
one just needs to fix the seed used to draw the samples.
With a bayesian analysis this becomes more complicated, and we advise to use combined fits as much as possible.

On the other hand, if the observables are uncorrelated (or obtained in a combined fit), then the fits can
be chained elegantly: for each observable we have one HMC trace which can be used as one component of some new data set on which the next
fit will be performed. 
Unlike a $\chi^2$ fit there is no assumption of gaussianity on this new dataset, it is left up to the model
to enforce it or not.
This could be important since there are many observables for which the gaussian approximation 
is not as good as it is for averaged primary observables.

\section{Variational inference}
We discussed how a bayesian model can be used with MLE or MAP, normal approximations around those, or a full Monte-Carlo sampling.
Variational inference (ADVI in PyMC) is an intermediate method we did not discuss, slightly simpler than the full sampling.
We did not obtain good results with this method in our exploration, but we did use it as a starting point for the thermalisation of our HMC.

The idea is to approximate the posterior with a family of base distributions. These base distributions can be anything, but uncorrelated
multivariate gaussians is a common choice. While this family of base distributions cannot generate the entire space of distributions,
this is sufficient for some uses. Instead of determining the posterior distribution as an arbitrary function, we are left with a few
parameters to optimise. This is done by minimising the Kullback-Leibler divergence.

This is different from a normal approximation around the MLE, something which can easily be seen by considering data distributed with a
bimodal distribution: the MLE chooses one peak while variational inference tends to cover both with a wider distribution.

\section{Binning and K folds}
\label{sec:binning}
If we cannot model the Monte-Carlo auto-correlations, a natural fallback technique is binning the data. This is still a possibility
with bayesian inference, just as it was for bootstraps or in principle even the $\Gamma$ method. As always, this leads to a simplification
of the modelling but at the cost of some loss of information in the data.

Somewhat similarly to binning, one can also fold the data, reshaping a size $n$ vector $y_i$ into a $\frac{n}{K}\times K$ matrix $y_{ij}$.
If our model describes this new data as $K$ identical independent copies using a common set of parameters, absolutely nothing changed.
However, applying LOO to the folded data actually results in a $K$-fold cross-validation on the original data, which could be more
appropriate to build an information criterion on auto-correlated data. Indeed, in general LOO does not apply to time series, and alternatives
such as Leave-Future-Out exist for these cases.

This folding also allows to reformulate the binning as a choice of model instead of a modification of the data: the unbinned version
has $K$ identical copies, so that for instance in the case of a gaussian likelihood this is like a $K$-component multivariate gaussian 
whose precision matrix is simply the identity, 
while a binned version can be built by choosing a rank-1 precision matrix.

\section{Over-shrinkage in MLE}
As stated in Sec.~\ref{sec:MLE}, shrinkage can improve the MLE. This can for instance be achieved through a Tikhonov regularisation (sometimes
improperly called a prior). However, one has to be careful when combining this with resampling.
Let us imagine a fit with a very flat direction, for instance the energy of a highly excited state in a correlator, or a direction which is only constrained by a unitarity inequality. Providing a weak
regularisation compared to our precision target is not sufficient to avoid the introduction of a bias: it has to be weaker than the information from the data. Otherwise for each bootstrap sample the maximum of the likelihood
would be very close to the {\it maximum} of the prior probability, and the bootstrap distribution could severely underestimate the uncertainty on the inferred parameter.
This can be cured by moving the prior for each bootstrap sample, but this problem simply does not happen with a bayesian inference.

\begin{acknowledgments} 
We thank Gregorio Herdo\'iza, Rainer Sommer, Carlos Pena and Lorenzo Barca for the discussions and comments on
the manuscript.
All computations in this work have been performed at the Deutsches Elektronen-Synchrotron (DESY).
We thank our colleagues in the CLS initiative for providing the ensemble on which our method is tested.
\end{acknowledgments}

\bibliography{bayesPaper}

\end{document}